\documentclass[twocolumn,showpacs,preprintnumbers,amsmath,amssymb]{revtex4}

\usepackage{graphicx}
\usepackage{dcolumn}
\usepackage{bm}

\begin{document}


\title{Quantitative Study of Magnetotransport through a (Ga,Mn)As Single Ferromagnetic Domain }

\author{S.~T.~B.~Goennenwein}\email{s.t.b.goennenwein@tnw.tudelft.nl}
\author{S.~Russo}
\author{A.~F.~Morpurgo}
\author{T.~M.~Klapwijk}
\affiliation{Kavli Institute of Nanoscience, Delft University of
Technology, Lorentzweg 1, 2628 CJ Delft, The Netherlands}
\author{W.~van Roy}
\author{J.~de Boeck}\altaffiliation[also at ]{Kavli Institute of Nanoscience, Delft University of
Technology, Lorentzweg 1, 2628 CJ Delft, The Netherlands}
\affiliation{IMEC, Kapeldreef 75, B-3001 Leuven, Belgium}

\date{\today}

\begin{abstract}
We have performed a systematic investigation of the longitudinal
and transverse magnetoresistance of a single ferromagnetic domain
in $\mathrm{Ga}_{1-x}\mathrm{Mn}_{x}\mathrm{As}$. We find that, by
taking into account the intrinsic dependence of the resistivity on
the magnetic induction, an excellent agreement between
experimental results and theoretical expectations is obtained. Our
findings provide a detailed and fully quantitative validation of
the theoretical description of magnetotransport through a single
ferromagnetic domain. Our analysis furthermore indicates the
relevance of magneto-impurity scattering as a mechanism for
magnetoresistance in
$\mathrm{Ga}_{1-x}\mathrm{Mn}_{x}\mathrm{As}$.
\end{abstract}


\maketitle

The resistance of ferromagnetic materials is a function of the
relative orientation between the magnetization $\mathbf{M}$ and
the current density $\mathbf{j}$. In general, the orientation of
$\mathbf{M}$ depends on the externally applied magnetic field,
which results in an anisotropic magnetoresistance (AMR)
characteristic of
ferromagnets\cite{McGuire:MTr-AMR:IEEETM:1975,Jan:review:1957}.
Although a vast amount of work has been devoted to the study of
this magnetization-induced AMR in the past, a complete
quantitative analysis of the magnetoresistive behavior of a single
ferromagnetic domain has not been performed yet.

It was found theoretically long ago from simple symmetry
considerations that, for isotropic materials, the AMR of a single
ferromagnetic domain can be described by the two
equations\cite{McGuire:MTr-AMR:IEEETM:1975,Jan:review:1957}
\begin{eqnarray}
  E_{\mathrm{long}}=j\rho_{\mathrm{long}}=j\rho_{\perp}+j\left( \rho_{\parallel}-\rho_{\perp}
  \right) \cos^2\phi_{M},\label{eq:MTr:sheet}\\
  E_{\mathrm{trans}}=j\rho_{\mathrm{trans}}=j\left( \rho_{\parallel}-\rho_{\perp}
  \right) \sin\phi_{M} \cos\phi_{M}\label{eq:MTr:Hall}.
\end{eqnarray}
Here, $E_{\mathrm{long}}$ and $E_{\mathrm{trans}}$ are the
components of the electric field along and perpendicular to
$\mathbf{j}$, $\rho_{\parallel}$ and $\rho_{\perp}$ are the value
of resistivity measured when $\mathbf{j}||\mathbf{M}$ and
$\mathbf{j}\perp\mathbf{M}$, and $\phi_{M}$ is the angle between
$\mathbf{M}$ and $\mathbf{j}$. These equations -- to which we will
refer to as the single domain magnetoresistance (SDM) model -- are
expected to account for the behavior of the longitudinal
\emph{and} the transverse voltage drop generated by a current
flowing through a ferromagnet. Given that the relative orientation
between $\mathbf{j}$ and $\mathbf{M}$ is known, the only
quantities that are needed to describe the (in general) complex
AMR behavior observed experimentally are $\rho_{\parallel}$ and
$\rho_{\perp}$.

In conventional metallic ferromagnets such as Ni or Co, a full
quantitative validation of the SDM model is difficult. Extensive
investigations of the longitudinal magnetoresistance have been
performed, but the relatively small magnitude of the transverse
(Hall-like) signal prevents a precise quantitative comparison of
experimental data to Eq.~(\ref{eq:MTr:Hall}). In the ferromagnetic
semiconductor $\mathrm{Ga}_{1-x}\mathrm{Mn}_{x}\mathrm{As}$, the
situation is different. A very large transverse electric field
("giant" planar Hall effect) has been recently
reported\cite{Tang:GaMnAs:GPHE:PRL90:2003} and its dependence on
the magnetization orientation found to be in excellent
quantitative agreement with the predictions of
Eq.~(\ref{eq:MTr:Hall}). However, the very rich behavior of the
longitudinal electric field has not been quantitatively analyzed
so far.

In this paper, we investigate magnetotransport through a single
domain of $\mathrm{Ga}_{1-x}\mathrm{Mn}_{x}\mathrm{As}$ to perform
a detailed and complete test of Eq.~(\ref{eq:MTr:sheet}) and
(\ref{eq:MTr:Hall}) of the SDM model. Our work is based on
measurements of the longitudinal and transverse magnetoresistance
for many different orientations of the applied in-plane magnetic
field. Specifically, we first analyze a selected set of these
measurements to determine the parameters of the SDM model,
including a linear dependence of $\rho_{\parallel}$ and
$\rho_{\perp}$ on the magnetic induction $\mathbf{B}$ (i.e.,
$\rho_{\perp},\rho_{\parallel}\propto-\mu_{0}|\mathbf{H}+\mathbf{M}|$).
We then use the values of the parameters so determined to perform
a fully quantitative comparison between the predictions of the SDM
model and magnetotransport measurements in arbitrary orientations
of the in-plane magnetic field. We find an excellent agreement
between theory and experiments in all cases, for \emph{both} the
longitudinal and the transverse magnetoresistance. The results
presented here conclusively demonstrate the full quantitative
validity of the SDM model, and confirm that the low-field AMR in
$\mathrm{Ga}_{1-x}\mathrm{Mn}_{x}\mathrm{As}$ is determined by one
single domain reversing its orientation via abrupt switches of
approximately $90^{\circ}$
\cite{Welp:GaMnAs:magnetization:PRL90:2003,Tang:GaMnAs:GPHE:PRL90:2003,Hamaya:GaMnAs:MTr:in-plane-ansiotropy-switching:JAP94:2003}.
In addition, our findings also indicate the relevance of
magneto-impurity scattering as a source of intrinsic
magnetoresistance in
$\mathrm{Ga}_{1-x}\mathrm{Mn}_{x}\mathrm{As}$.

\begin{figure}
\includegraphics[width=8cm]{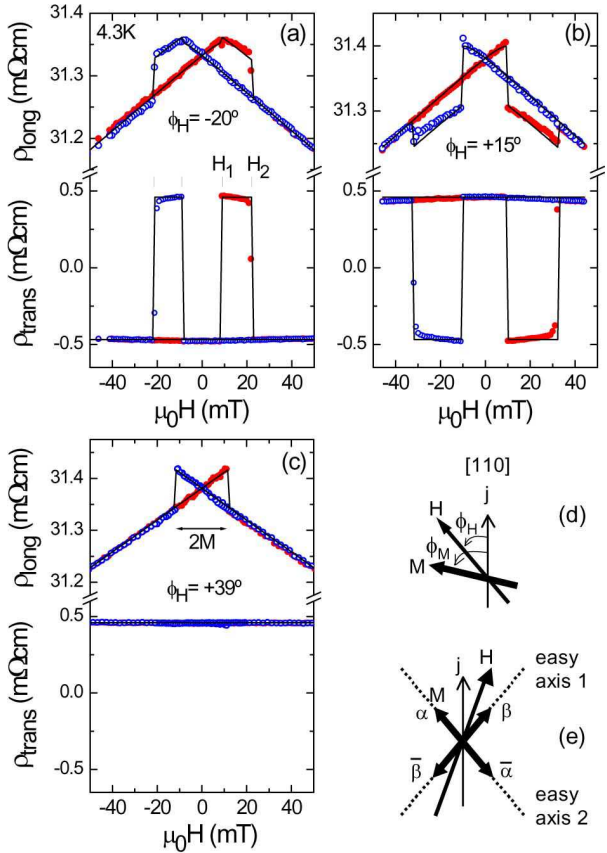}
\caption{Panels (a--c) show the longitudinal and transverse
magnetoresistance of a single
$\mathrm{Ga}_{1-x}\mathrm{Mn}_{x}\mathrm{As}$ ferromagnetic
domain, measured for three different orientations of the in-plane
magnetic field ($\phi_H = -20^\circ, +15^\circ$, and $+39^\circ$,
respectively). In all panels, full symbols represent the data
measured for an increasing magnetic field $H$ (up-sweep) and the
open symbols those measured with a decreasing field (down-sweep).
The full lines represent the resistances calculated from
Eqs.~(\ref{eq:MTr:sheet}) and (\ref{eq:MTr:Hall}), as discussed in
the text. Panel (d) defines the angles $\phi_{H}$ and $\phi_{M}$
between the field $\mathbf{H}$, the magnetization $\mathbf{M}$,
and the current density $\mathbf{j}$. Panel (e) illustrates the
orientation of the two easy axes in
$\mathrm{Ga}_{1-x}\mathrm{Mn}_{x}\mathrm{As}$ with respect to the
direction of the current flow. When the field $H$ is swept,
magnetization reversal occurs by subsequent,
$\approx90^\circ$-switches, e.g.~$\beta \rightarrow
\overline{\alpha} \rightarrow
\overline{\beta}$.}\label{fig:MTr-fig1-twotraces}
\end{figure}

The $\mathrm{Ga}_{1-x}\mathrm{Mn}_{x}\mathrm{As}$ thin films
studied were grown by molecular beam epitaxy on (100)-oriented,
semi-insulating GaAs wafers. Here, we will focus on a sample with
Mn content $x=0.07$, thickness $d=57~\mathrm{nm}$, and Curie
temperature $T_{\mathrm{C}}\approx80~\mathrm{K}$. The thin film
was patterned into $200\times 50~\mathrm{\mu m}^2$ Hall bar mesas
using photolithography and wet chemical etching, with the long
axes of the Hall bars aligned along the crystallographic $[110]$
direction to within experimental accuracy. For the
magnetotransport experiments, the samples were mounted on a
rotatable sample holder and inserted into the He cryostat of a
superconducting magnet. $\rho_{\mathrm{long}}$ and
$\rho_{\mathrm{trans}}$ were simultaneously recorded in ac
current-bias 4-probe measurements, using a lock-in technique. In
all experiments discussed here, the external magnetic field
$\mathbf{H}$ was applied in the plane of the thin film. The
rotatable holder enabled us to choose the angle $\phi_{H}$ between
$\mathbf{j}$ and $\mathbf{H}$ (see
Fig.~\ref{fig:MTr-fig1-twotraces}(d)) with an accuracy of about
$1^\circ$.

\begin{figure}
\includegraphics[width=7cm]{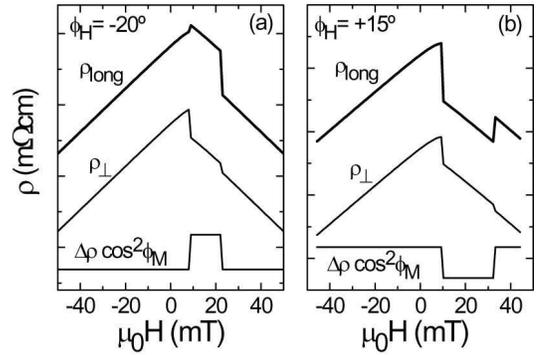}
\caption{Panel (a) and (b) illustrate the behavior of the
different contributions to the longitudinal magnetoresistance
(thin lines) as well as their sum (i.e., the total longitudinal
resistance; thick line), as calculated from
Eq.~(\ref{eq:MTr:sheet}) for two different orientations of the
in-plane field ($\phi_H= -20^{\circ}$ and $+15^{\circ}$). All the
traces are offset for clarity. The first term, $\rho_{\perp}$, is
proportional to the magnitude of the magnetic induction vector.
The second term, $\left( \rho_{\parallel}-\rho_{\perp}\right)
\cos^2(\phi_M)=\Delta \rho \cos^2(\phi_M) $, reflects the abrupt
switches of $\phi_{M}$. A comparison between calculations and
experimental data is shown in
Fig.~\ref{fig:MTr-fig1-twotraces}(a,b) and in
Fig.~\ref{fig:MTr-Fig3-lotsoftraces}.}\label{fig:MTr-fig2-simu}
\end{figure}

The rich features in the low-field, anisotropic magnetoresistance
of $\mathrm{Ga}_{1-x}\mathrm{Mn}_{x}\mathrm{As}$ are illustrated
in Figure \ref{fig:MTr-fig1-twotraces}, which shows the
magnetoresistance measured for three different orientations of the
in-plane magnetic field. It is apparent that the shape of the
observed magnetoresistance, which is in general hysteretic as
expected for a ferromagnet, strongly depends on the field
direction. Whereas $\rho_{\mathrm{trans}}$ only switches between
two different and approximately constant levels (of $\simeq \pm
0.45$ m$\Omega$cm), $\rho_{\mathrm{long}}$ exhibits a more complex
behavior. Note, in particular, that even for those field
orientations in which $\rho_{\mathrm{trans}}$ does not show any
dependence on $H$ ($\phi_{H}=+39^\circ$,
Fig.~\ref{fig:MTr-fig1-twotraces}(c)), a strong field dependence
as well as hysteresis are still present in $\rho_{\mathrm{long}}$.
While similar observations have already been reported, no full
quantitative analysis has been performed: the complexity of the
behavior of the longitudinal magnetoresistance demonstrates that
for the validation of the SDM model, a careful test of both
Eq.~(\ref{eq:MTr:sheet}) and (\ref{eq:MTr:Hall}) is needed.

The analysis of the transverse magnetoresistance signal and the
comparison to Eq.~(\ref{eq:MTr:Hall}) is identical to what has
been discussed by Tang {\it et al.} in
Ref.~\cite{Tang:GaMnAs:GPHE:PRL90:2003}, to which we refer the
reader for details. Here, we simply summarize the results which
are needed in the analysis of the longitudinal
magnetoresistance.\cite{Welp:GaMnAs:magnetization:PRL90:2003,Tang:GaMnAs:GPHE:PRL90:2003,Hamaya:GaMnAs:MTr:in-plane-ansiotropy-switching:JAP94:2003,Cowburn:MTR-switching:PRL:1997,Tang:GaMnAs:domain-wall-MR:Nature:2004}
In particular, the transverse in-plane magnetoresistance and its
dependence on the orientation of the external magnetic field can
be completely understood in terms of abrupt changes of the
magnetization orientation in the
$\mathrm{Ga}_{1-x}\mathrm{Mn}_{x}\mathrm{As}$ layer. At low
temperature and in the magnetic field range used in our
experiments, the magnetization is always parallel or anti-parallel
to one of the two easy magnetic axes. These axes point
approximately along the $[100]$ and $[010]$ directions
(Fig.~\ref{fig:MTr-fig1-twotraces}(e)). When the magnetic field is
swept, magnetization reorientation occurs via the nucleation and
rapid expansion of a $90^\circ$
domain.\cite{Welp:GaMnAs:magnetization:PRL90:2003} On the time
scale of magnetotransport experiments, this process appears as an
abrupt switch of the direction of $\mathbf{M}$. It takes place
when the energy gained by the magnetization reorientation becomes
larger than a fixed domain wall pinning energy. From the analysis
of the transverse magnetoresistance in our samples, we find that
the orientations of the two easy axes correspond to
$\phi_{H}=+39^\circ$ (direction $\alpha$ in
Fig.~\ref{fig:MTr-fig1-twotraces}(d)) and $\phi_{H}=-42^\circ$
(direction $\beta$). The fields $H_1$ and $H_2$, at which the
magnetization switches, are quantitatively consistent with the
above magnetization reorientation picture for all field
orientations $\phi_H$. Finally, from the analysis of
$\rho_{\mathrm{trans}}$ we obtain that
$\rho_{\parallel}-\rho_{\perp}=0.96~\mathrm{m\Omega cm}$,
independent of $H$. All these results fully agree with the
findings of Ref. \cite{Tang:GaMnAs:GPHE:PRL90:2003} and validate
Eq.~(\ref{eq:MTr:Hall}).

We now proceed to the analysis of $\rho_{\mathrm{long}}$. We start
by considering the simplest case, in which the applied magnetic
field is oriented along one of the easy axes (e.g., $\phi_H
=+39^\circ$; see Fig.~\ref{fig:MTr-fig1-twotraces}(c)). In this
case, upon sweeping the external magnetic field, the magnetization
simply reverts its direction ($H_1=H_2$, and $\phi_M$ switches
from $+39^\circ$ to $(39+180)^\circ$). As a consequence, the
$\cos^2(\phi_M)$ term in Eq.~(\ref{eq:MTr:sheet}), as well as the
$\sin(\phi_M)\cos(\phi_M)$ term in Eq.~(\ref{eq:MTr:Hall}), are
constant upon magnetization reversal. This is indeed observed
experimentally in the behavior of $\rho_{\mathrm{trans}}$, which
is exactly constant for this particular orientation of the
external field (see Fig.~\ref{fig:MTr-fig1-twotraces}(c)).
Nevertheless, it is apparent from
Fig.~\ref{fig:MTr-fig1-twotraces}(c) that a linear
magnetoresistance, as well as resistance jumps at the magnetic
field values for which $\mathbf{M}$ reverses, still are present in
$\rho_{\mathrm{long}}$. Since the second term on the right-hand
side of Eq.~(\ref{eq:MTr:sheet}) does not vary with $H$, we
conclude that the linear magnetoresistance and the jumps observed
experimentally in $\rho_{\mathrm{long}}$ originate from a magnetic
field dependence of $\rho_{\perp}$, i.e., the first term on the
right-hand side of Eq.~(\ref{eq:MTr:sheet}).

The observed behavior can be understood if $\rho_{\perp}$ is a
function of $B= |\mu_0(\mathbf{H}+\mathbf{M})|$, since then the
reversal of $\mathbf{M}$ results in an abrupt change of $B$ and,
consequently, of $\rho_{\perp}$. To reproduce the experimental
data, we take $\rho_{\perp}(B) = a + b \times B$ with
$a=30.85~\mathrm{m}\Omega\mathrm{cm}$ and
$b=-3.06~\mathrm{m}\Omega\mathrm{cm}/\mathrm{T}$. The continuous
line superimposed on top of the $\rho_{\mathrm{long}}$ data in
Fig.~\ref{fig:MTr-fig1-twotraces}(c) illustrates the precision
with which the chosen $\rho_{\perp}(B)$ reproduces the
experimental data for $\phi_H=+39^\circ$. In what follows, we will
use this dependence of $\rho_{\perp}$ on $B$ to analyze the
longitudinal magnetoresistance for all the other field
orientations, with {\it one and the same value of the parameters
$a$ and $b$ given above}. The microscopic origin of the
$B$-dependence of $\rho_{\perp}$ will be discussed at the end.

We also note that the linear dependence of $\rho_{\perp}$ on $B$
allows us to extract the magnitude of $\mathbf{M}$ in our samples.
In particular, it follows from the linear dependence of
$\rho_{\perp}$ on $B$ that the difference in $H$ between the two
jumps in $\rho_{\mathrm{long}}$ corresponds to $2 M$
(Fig.~\ref{fig:MTr-fig1-twotraces}(c)). The magnetization thus
obtained is $\mu_0 M = 12~\mathrm{mT}$, which is somewhat smaller
than the saturation magnetization that one would expect from the
nominal Mn content $x=0.07$ assuming that all the spins are
aligned. Such a magnetization deficit is often observed in
$\mathrm{Ga}_{1-x}\mathrm{Mn}_{x}\mathrm{As}$.\cite{vanEsch:MinGaMnAs:PRB:1997,Korzhavyi:As-defect-and-magnetism:GaMnAS:PRL88:2002,Wang:GaMnAs:annealing:JAP:2004}

The analysis of $\rho_{\perp}$ for $\phi_H = +39^\circ$ and of
$\rho_{\mathrm{trans}}$ completely determines all quantities
entering Eq.~(\ref{eq:MTr:sheet}) and (\ref{eq:MTr:Hall}). We thus
can now check if the SDM model reproduces the magnetoresistance
behavior observed experimentally for arbitrary orientations of the
applied in-plane magnetic field. We emphasize that at this point,
there are no adjustable parameters: we just use the SDM model to
calculate the magnetoresistance and see if the theoretical curves
match the experimental ones. To illustrate how the theoretical
curves are calculated, we have plotted the different terms on the
right-hand side of Eq.~(\ref{eq:MTr:sheet}) separately in
Fig.~\ref{fig:MTr-fig2-simu} for two specific orientations
$\phi_{H}$ of the external magnetic field.

\begin{figure}
\includegraphics[width=7cm]{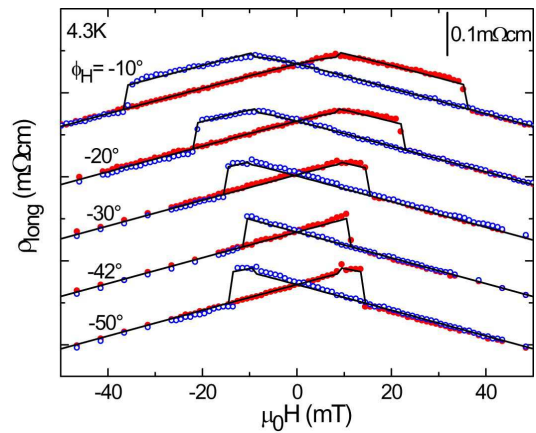}
\caption{\label{fig:MTr-Fig3-lotsoftraces} Comparison between the
longitudinal magnetoresistance measured for different orientations
$\phi_{H}$ of the magnetic field (symbols) and the predictions by
Eq.~(\ref{eq:MTr:sheet}) of the SDM model (full lines) in which
all parameters have been fixed by our analysis. The
magnetoresistance traces have been offset for clarity. Full
symbols represent the experimental data measured for magnetic
field up-sweeps, open symbols correspond to the data measured for
magnetic field down-sweeps.}
\end{figure}

\begin{figure}
\includegraphics[width=8cm]{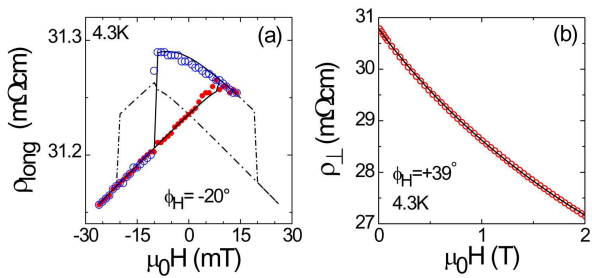}
\caption{\label{fig:MTr-Fig4-minorloops} (a) Longitudinal
magnetoresistance measured ramping up (full circles) and down
(open circles) the magnetic field, reverting the sweep direction
at $15~\mathrm{mT}$, i.e., in between the two switches of the
magnetization orientation. The full line is calculated
theoretically using Eq.~(\ref{eq:MTr:sheet}) of the SDM model,
without any adjustable parameters (all parameters are fixed by the
previous analysis). The dash-dotted line illustrates the
magnetoresistance obtained if the field sweep direction is
reversed after the second switch of the magnetization orientation
(see also Fig.~\ref{fig:MTr-fig1-twotraces}(a)). (b) The magnetic
field dependence of $\rho_{\perp}$ measured experimentally (open
symbols) is reproduced by the magneto-impurity scattering model
prosed by Nagaev (full line) over a broad range of magnetic
fields.}
\end{figure}

The behavior of the $\cos^2\phi_{M}$-term is completely analogous
to the $\sin\phi_{M}\cos\phi_{M}$-term in Eq.~(\ref{eq:MTr:Hall}).
It reflects the abrupt changes in the magnetization orientation
$\phi_{M}$ as the field is swept. When the magnetization switches
from one easy axis to the other at the fields $H_1$ and $H_2$, the
magnitude of $\left( \rho_{\parallel}-\rho_{\perp} \right)
\cos^2\phi_{M}$ abruptly changes, resulting in square-like steps.
The behavior of the first term, $\rho_{\perp}(B)$, is different.
Because $\rho_{\perp}(B)\propto |\mathbf{H}+\mathbf{M}|$, this
term retraces the magnitude of the magnetic induction vector. The
abrupt reorientations of $\mathbf{M}$ also lead to abrupt steps in
$|\mathbf{H}+\mathbf{M}|$, the magnitude of which is different for
different orientations of $\mathbf{H}$, due to the vectorial
addition of $\mathbf{M}$ and $\mathbf{H}$. Because the magnitude
of $M$ and $H$ as well as their orientations are known,
$|\mathbf{H}+\mathbf{M}|$ and thus also $\rho_{\perp}(B)$, can be
calculated. Taken together, the interplay of the two terms on the
right side of Eq.~(\ref{eq:MTr:sheet}) results in the
qualitatively different shape of $\rho_{\mathrm{long}}$ for
different orientations of the magnetic field
(Fig.~\ref{fig:MTr-fig2-simu}).

The $\rho_{\mathrm{long}}$ and $\rho_{\mathrm{trans}}$ calculated
from the SDM model are shown as full lines in
Figs.~\ref{fig:MTr-fig1-twotraces}(a,b) and
\ref{fig:MTr-Fig3-lotsoftraces}. For all orientations $\phi_{H}$
of the external magnetic field, the SDM model quantitatively
describes the longitudinal resistance observed in experiment. In
particular, both the strongly varying magnitude and shape of the
switches observed in $\rho_{\mathrm{long}}$ are precisely
reproduced in the calculations. This demonstrates that
Eqs.~(\ref{eq:MTr:sheet}) and (\ref{eq:MTr:Hall}) yield a fully
quantitative description of the low-field AMR of
$\mathrm{Ga}_{1-x}\mathrm{Mn}_{x}\mathrm{As}$ for all orientations
$\phi_{H}$ of the external magnetic field.

The agreement between the SDM model and the experiment can be
tested even further by analyzing magnetoresistance measurements in
which the direction of the magnetic field sweep is reversed after
the first, but before the second magnetization switch. In this
case, for small values of $H$, $\mathbf{M}$ points along one easy
axis in the up sweep and along the other in the down sweep. As a
consequence, the magnetoresistance traces for the up and the down
sweep are different, as illustrated in
Fig.~\ref{fig:MTr-Fig4-minorloops}(a). Again, an excellent
agreement between the SDM model and experiment is found, with no
adjustable parameters. This agreement confirms our hypothesis that
$\rho_{\perp}(B)\propto -|\mathbf{B}|$, since around $H=0$, the
magnetoresistance only originates from a change in the vectorial
sum $\mu_{0}\left( \mathbf{H}+\mathbf{M}\right)$.

We now briefly discuss the microscopic origin of the
experimentally observed $B$-dependence of $\rho_{\perp}$ (as well
as of $\rho_{\parallel}$, since
$\rho_{\perp}-\rho_{\parallel}=\mathrm{const.}$). Although several
different mechanisms, such as weak localization, spin-disorder
scattering, and variable-range hopping give rise to a magnetic
field dependent
resistivity,\cite{Matsukura:GaMnAs:MTR:sheet:inplane:PE21:2004,Yoon:GaMnAs:MTr:JAP:2004,Dietl-MTr-condmat-revised,Matsukura:GaMnAsRKKYandMncontent:PRB:1998,vanEsch:MinGaMnAs:PRB:1997}
none of them predicts a linear negative magnetoresistance.
However, it was found recently by
Nagaev\cite{Nagaev:review:MTR:PhysicsReports:2001,Nagaev:Magnetoimpurity:PRB58:1998}
that spin-dependent scattering at magnetic impurities can result
in a negative magnetoresistance of the form
$\rho(B)=a-bB+cB^{3/2}$. This functional dependence correctly
describes our experimental data not only in the linear regime that
we have discussed here extensively, but also for higher field
values, where a non-linear dependence is experimentally visible
(see Fig.~\ref{fig:MTr-Fig4-minorloops}(b)). This observation, in
conjunction with recent investigations of the resistance
temperature dependence\cite{Yuldashev:GaMnAs:MTR:APL2003},
indicates the relevance of magneto-impurity scattering in
$\mathrm{Ga}_{1-x}\mathrm{Mn}_{x}\mathrm{As}$. In the future, a
systematic comparison between experimental data and the
magneto-impurity model will be needed to establish up to which
extent this theory describes the intrinsic magnetoresistance of
$\mathrm{Ga}_{1-x}\mathrm{Mn}_{x}\mathrm{As}$. The quantitative
understanding of anisotropic magnetoresistance reached in our work
makes it possible to separate the magnetization-related AMR from
the intrinsic magnetoresistance ($\rho_{\parallel},\rho_{\perp}$),
thus enabling a detailed study of the latter.

In summary, we have investigated experimentally the longitudinal
and transverse magnetoresistance in a single ferromagnetic domain
of $\mathrm{Ga}_{1-x}\mathrm{Mn}_{x}\mathrm{As}$ and performed a
comprehensive theoretical analysis of the experimental results. We
find that, by taking into account the intrinsic dependence of the
resistivity on the magnetic induction, excellent quantitative
agreement between experiments and theory is obtained. Our results
provide the first detailed and fully quantitative validation of
the theory for magnetotransport through a single ferromagnetic
domain. In addition, our analysis indicates the relevance of
magneto-impurity scattering in
$\mathrm{Ga}_{1-x}\mathrm{Mn}_{x}\mathrm{As}$.

The authors gratefully acknowledge stimulating discussions with
G.~E.~W.~Bauer and R.~Gross. The work of A.F.M. and of S.T.B.G. is
part of the NWO Vernieuwingsimpuls 2000 program.

\end{document}